\documentclass[12pt]{article}
\usepackage{amssymb}
\begin{document}

\title{\bf Massive Relativistic Particle Models \\ with
Bosonic Counterpart \\ of Supersymmetry}

\author{ {\bf Sergey Fedoruk}\thanks{On leave of absence from
Ukrainian Engineering--Pedagogical Academy, Kharkov, Ukraine.} \ and
\ {\bf Jerzy
 Lukierski}\thanks{Supported by KBN grant 1P03B01828.}
 \\
  Institute for Theoretical Physics, \\
  University of Wroc{\l}aw, pl.
Maxa Borna 9, 50-204 Wroc{\l}aw, Poland}

\date{}
\maketitle

\begin{abstract} We consider the massive relativistic particle models  on
fourdimensional Minkowski space extended by $N$ commuting Weyl spinors
for $N=1$ and $N=2$. The $N=1$ model is invariant under the most
general form of bosonic counterpart of simple $D=4$ supersymmetry,
and provides after quantization the bosonic counterpart of chiral
superfields, satisfying Klein--Gordon equation.
 In massless case these fields do satisfy the Fierz-Pauli
 equations.
 For $N=2$ we obtain after
quantization the free massive higher spin fields for arbitrary spin
satisfying linear Bargman--Wigner equations. Finally the problem of
 statistics  in presented framework for half--integer classical spin fields is discussed.
\end{abstract}
\thispagestyle{empty}

\newpage

\setcounter{page}{1}
\section{Introduction}

Important extensions of the relativistic symmetries were considered in
the following two directions:
\begin{enumerate}
    \item Supersymmetric extension, relating by supersymmetry (SUSY)
    transformations integer and half--integer spin fields
    (see e. g.~\cite{Wess, GGRS}). The geometric
    way of describing supersymmetric multiplets is realized
    in terms of superfields -- the functions on superspace
    $Y_A =(x_\mu, \theta_\alpha^i, \bar\theta_{\dot\alpha}^i)$ where
    $\theta_\alpha^i$ are anticommuting Grassmann spinors.
\item Introduction of higher spin (HS) algebras, which act on
    infinite spin multiplets or if $m=0$ on infinite helicity multiplets
    (see e. g.~\cite{Vas1}-\cite{Sor}).
The
    representation spaces of HS algebras are described by the functions on `bosonic'
    superspace $Z_A=(x_\mu, \lambda_\alpha^i,
    \bar\lambda_{\dot\alpha}^i)$ with
    additional commuting spinor variables $\lambda_\alpha^i$.
The bosonic counterparts of superfields
    one can call the  spinorial
    Kaluza--Klein (KK) fields, with spinorial additional dimensions. We obtain
    \begin{equation}
    \Phi_A(Z_A)=\sum_{n,k=0}^\infty
    \sum_{{}^{(\alpha_1\dots\alpha_n)}_{(\dot\beta_1\dots\dot\beta_k)}}
    \sum_{{}^{(i_1\dots i_n)}_{(j_1\dots j_k)}}\varphi^{\alpha_1\dots\alpha_n
    \dot\beta_1\dots\dot\beta_k}_{A;i_1\dots
    i_n j_1\dots j_k}(x)
    \lambda_{\alpha_1}^{i_1}\dots\lambda_{\alpha_n}^{i_n}
    \bar\lambda_{\dot\beta_1}^{j_1}\dots\bar\lambda_{\dot\beta_k}^{j_k}\,.
    \end{equation}
\end{enumerate}

The auxiliary commuting spinorial variables $(\lambda_\alpha^i,
\bar\lambda_{\dot\alpha}^i)$ occurs in several geometric frameworks,
 for example in twistor approach to the space--time
geometry~\cite{PenMac}-\cite{Bet} or in the models with double
(target and world volume) supersymmetry~\cite{KHAL}-\cite{Sorok}.

In this note we would like to study the group-theoretic and
dynamical consequences of introducing bosonic counterpart of
supersymmetry, obtained by supplementing the Poincar\'{e} algebra by bosonic
spinorial charges. We recall the general $N=1$ SUSY relation with
tensorial  charges~\cite{FerPor,BandLuk}
\begin{equation}
\{Q_a, Q_b \}= 2(\gamma^\mu C)_{ab}P_\mu
+(\sigma^{\mu\nu}C)_{ab}Z_{\mu\nu}
\end{equation}
where in Majorana representation $C=\gamma_0$ and
\begin{description}
    \item -- $Q_a$ is a four--component Majorana spinor of
    supercharges
    \item -- $Z_{\mu\nu}=-Z_{\nu\mu}$ describe six Abelian tensorial
    charges.
\end{description}
The bosonic counterpart of general $N=1$ SUSY takes the form
\begin{equation}\label{bsusy}
[R_a, R_b ]= 2(\gamma^\mu \gamma_5 C)_{ab} P_\mu +2C_{ab} Z^{(1)}
+ 2(\gamma_5 C)_{ab} Z^{(2)}
\end{equation}
where
\begin{description}
    \item -- $R_a$ is a four--component spinor of
    bosonic charges
    \item -- $Z^{(1)}$ ($Z^{(2)}$) are scalar (pseudoscalar)
    central charges.
\end{description}
In order to obtain in~(\ref{bsusy}) the standard inversion
properties of the fourmomentum generator one should assume suitable
transformation properties of the spinor $R_a$.\footnote{Spinorial
supercharges transform under space--time inversions in standard way
($Q_a^\prime =(\gamma_0 Q)_a$ for the space inversion $P$,
$Q_a^\prime =(\gamma_0\gamma_5 Q)_a$ for the time inversion $T$).
The bosonic spinorial charges $R_a$ are so--called
pseudospinor~\cite{Rzew, Bog}) transforming under inversion in
alternative way ($R_a^\prime =(\gamma_0\gamma_5 R)_a$ under $P$,
$R_a^\prime =(\gamma_0 R)_a$ under $T$).}

Our aim is to study the massive relativistic particle models
invariant under bosonic counterpart of SUSY and perform their
quantization. Contrary to the case of simple SUSY the $N=1$
relation~(\ref{bsusy}) contains scalar and pseudoscalar  central charges,
which  can
be related with the mass parameter. In Sect. 2 we describe (using
two--component Weyl notation) the particle model describing the
trajectory in the spinorial KK space ${\cal M}^{4,4}$ with the
coordinates $Z_A=(x_\mu, \lambda_\alpha, \bar\lambda_{\dot\alpha})$.
After calculating the complete set of constraints we perform the
quantization using either Heisenberg picture or the
  Gupta--Bleuler method (Schr\"{o}dinger picture).\footnote{Gupta--Bleuler
method has been applied to massive relativistic superparticle e.
g. in~\cite{AzLuk, LusMil}.} We shall obtain the   wave function
$\Psi(Z_A)$ satisfying the KG equation and the bosonic counterpart
of the chirality condition. In Sect. 3 we analyze the massless
limit of our model, with massless fields with arbitrary helicity
satisfying Fierz-Pauli equations.
 In Sect. 4 we consider the
relativistic particle in $N=2$ spinorial KK space ${\cal M}^{4,8}$
with the coordinates $(x_\mu, \lambda_{\alpha i},
\bar\lambda_{\dot\alpha i})$ $(i=1,2)$. It appears that for the
particular choice of bosonic counterpart of $N=2$ SUSY, with
internal symmetry $O(1,1)$,
  one can obtain the linear Bargman--Wigner equations
for $D=4$ massive higher spin fields~\cite{BargWig,Bog}. In Sect.
5 we shall discuss the problem of  nonstandard relation between
spin and statistics for the field components of spinorial KK
fields.

\section{Massive particle model with $N=1$ bosonic counterpart of SUSY.}
\subsection{Classical model}
We consider the following action\footnote{We use following
notations. The metric has mostly minus $\eta_{\mu\nu}={\rm
diag}(+---)$. The Weyl two--spinor indices are risen and lowered by
$\varphi^\alpha=\epsilon^{\alpha\beta}\varphi_\beta$,
$\varphi_\alpha=\varphi^\beta\epsilon_{\beta\alpha}$,
$\bar\varphi^{\dot\alpha}=\epsilon^{\dot\alpha\dot\beta}\bar\varphi_{\dot\beta}$,
$\bar\varphi_{\dot\alpha}=\bar\varphi^{\dot\beta}\epsilon_{\dot\beta\dot\alpha}$
where
$\epsilon^{\alpha\beta}\epsilon_{\beta\gamma}=-\delta^{\alpha}_{\gamma}$,
$\epsilon^{\dot\alpha\dot\beta}\epsilon_{\dot\beta\dot\gamma}=
-\delta^{\dot\alpha}_{\dot\gamma}$. Algebra $\sigma$--matrices
$\sigma^\mu_{\alpha\dot\beta}=(\overline{\sigma^\mu_{\beta\dot\alpha}})$
and $\sigma_\mu^{\dot\alpha\alpha}=\epsilon^{\alpha\beta}
\epsilon^{\dot\alpha\dot\beta}\sigma_{\mu\beta\dot\beta}$ is
$\sigma_{\mu\alpha\dot\gamma}\sigma_\nu^{\dot\gamma\beta}+
\sigma_{\nu\alpha\dot\gamma}\sigma_\mu^{\dot\gamma\beta}=
2\eta_{\mu\nu}\delta_{\alpha}^{\beta}$. Also we define
$p_{\alpha\dot\beta}=p_{\mu}\sigma^\mu_{\alpha\dot\beta}$,
$p^{\dot\alpha\beta}=p^{\mu}\sigma_\mu^{\dot\alpha\beta}$ for any
vector $p_{\mu}$.}
\begin{equation}\label{act}
S=\int d\tau\,{\cal L} \,,
\end{equation}
\begin{equation}\label{Lagr1}
{\cal L}=-m(\dot\omega_\mu\dot\omega^\mu)^{1/2} -i
(z\dot\lambda^\alpha\lambda_\alpha - \bar
z\bar\lambda_{\dot\alpha}\dot{\bar\lambda}{}^{\dot\alpha})
\end{equation}
where
\begin{equation}\label{om}
d\omega^\mu=\dot\omega^\mu d\tau=dx^\mu-
id\lambda^\alpha\sigma^\mu_{\alpha\dot\beta}\bar\lambda^{\dot\beta}+
i\lambda^\alpha\sigma^\mu_{\alpha\dot\beta}d\bar\lambda^{\dot\beta}\,.
\end{equation}
The action~(\ref{act})-(\ref{om}) describes the particle trajectory
in Minkowski space extended by two commuting complex Weyl spinor
coordinates $\lambda^\alpha(\tau)$,
$\bar\lambda^{\dot\alpha}=(\overline{\lambda^\alpha})$ and invariant
under the following spinorial bosonic transformation
\begin{equation}\label{trans}
\delta x^\mu=
i\lambda^\alpha\sigma^\mu_{\alpha\dot\beta}\bar\varepsilon^{\dot\beta}-
i\varepsilon^\alpha\sigma^\mu_{\alpha\dot\beta}\bar\lambda^{\dot\beta}\,,
\qquad \delta\lambda^\alpha=\varepsilon^\alpha \,, \qquad
\delta\bar\lambda^{\dot\alpha}=\bar\varepsilon^{\dot\alpha}
\end{equation}
where $\varepsilon^\alpha$ is a constant commuting Weyl spinor.
The constant $m$ is the mass of particle whereas $z$ is an arbitrary
complex parameter with the dimension of mass. It is easy to see that
 performing the suitable phase transformation $\lambda'_{\alpha}= e^{i a}\lambda_{\alpha}$,
 $\bar{\lambda}'_{\dot{\alpha}}=e^{-i a}\bar{\lambda}_{\dot{\alpha}}$, where
  $a= \frac{1}{2} arg \, z$ one gets the real parameter $z$.

Conserved Noether spinorial charges corresponding to the
transformations~(\ref{trans}) are
\begin{equation}\label{R1}
R_\alpha \equiv \pi_\alpha
-ip_{\alpha\dot\beta}\bar\lambda^{\dot\beta}-i z\lambda_\alpha
\,,
\end{equation}
\begin{equation}\label{bR1}
\bar R_{\dot\alpha}\equiv \bar\pi_{\dot\alpha} + i\lambda^\beta
p_{\beta\dot\alpha} +i\bar z\bar\lambda_{\dot\alpha}
\end{equation}
where the canonical momenta are defined by
\begin{equation}\label{px1}
p_\mu=\frac{\partial {\cal L}}{\partial\dot x^\mu}=
-m(\dot\omega_\nu\dot\omega^\nu)^{-1/2}\dot\omega_\mu\,,
\end{equation}
\begin{equation}\label{pi1}
\pi_\alpha =\frac{\partial {\cal L}}{\partial\dot\lambda^\alpha}=
-ip_{\alpha\dot\beta}\bar\lambda^{\dot\beta}-iz\lambda_\alpha\,,
\end{equation}
\begin{equation}\label{bpi1}
\bar\pi_{\dot\alpha} =\frac{\partial {\cal
L}}{\partial\dot{\bar\lambda}{}^{\dot\alpha}}= i\lambda^\beta
p_{\beta\dot\alpha} +i\bar z\bar\lambda_{\dot\alpha}\,.
\end{equation}
Using the canonical Poisson brackets
\begin{equation}\label{CCR}
\{x^\mu, p_\nu\}=\delta^\mu_\nu\,, \qquad \{\lambda^\alpha,
\pi_\beta\}=\delta^\alpha_\beta\,, \qquad
\{{\bar\lambda}^{\dot\alpha},\bar\pi_{\dot\beta}\}=
\delta^{\dot\alpha}_{\dot\beta}
\end{equation}
we obtain the algebra
\begin{equation}\label{al1-ch}
\{R_\alpha, \bar R_{\dot\beta}\}=-2ip_{\alpha\dot\beta}\,,
\end{equation}
\begin{equation}\label{al2-ch}
\{R_{\alpha}, R_{\beta}\}=2i z\epsilon_{\alpha\beta}\,,\qquad
\{\bar R_{\dot\alpha}, \bar R_{\dot\beta}\}= -2i\bar z
\epsilon_{\dot\alpha\dot\beta}
\end{equation}
which is equivalent to the algebra~(\ref{bsusy}) with
$Z=Z^{(1)}+iZ^{(2)}=z$.

From~(\ref{px1})--(\ref{bpi1}) follow the mass shell constraint and the set of four
spinorial constraints
\begin{equation}\label{T1}
T \equiv p^2 -m^2\approx 0\,,
\end{equation}
\begin{equation}\label{D1}
D_\alpha \equiv \pi_\alpha
+ip_{\alpha\dot\beta}\bar\lambda^{\dot\beta}+i
z\lambda_\alpha\approx 0\,,
\end{equation}
\begin{equation}\label{bD1}
\bar D_{\dot\alpha}\equiv \bar\pi_{\dot\alpha} - i\lambda^\beta
p_{\beta\dot\alpha} -i\bar z\bar\lambda_{\dot\alpha}\approx 0\,
\end{equation}

Using the formulae~(\ref{px1})-(\ref{bpi1}) we confirm that the
canonical Hamiltonian vanishes\footnote{The vanishing of Hamiltonian follows from
the invariance of the action
(\ref{act}--\ref{om})
 under the arbitrary local rescaling
 $\tau \to \tau'= \tau'(\tau)$.}
$$
{\cal H}=\dot x^\mu p_\mu+ \dot\lambda^\alpha \pi_\alpha+
\bar\pi_{\dot\alpha} \dot{\bar\lambda}{}^{\dot\alpha}-{\cal
L}=0\,.
$$
and the total Hamiltonian is the linear combination of first class
constraints multiplied by Lagrange multipliers.

The constraints~(\ref{T1})-(\ref{bD1}) satisfy the following Poisson
brackets
\begin{equation}\label{al1}
\{D_\alpha, \bar D_{\dot\beta}\}=2ip_{\alpha\dot\beta}\,,
\end{equation}
\begin{equation}\label{al2}
\{D_{\alpha}, D_{\beta}\}=-2i z\epsilon_{\alpha\beta}\,,\qquad
\{\bar D_{\dot\alpha}, \bar D_{\dot\beta}\}= 2i \bar z
\epsilon_{\dot\alpha\dot\beta}\,.
\end{equation}
The scalar constraint $T\approx 0$ is first class and all the
spinorial constraints~(\ref{D1}), (\ref{bD1}) are second class.
Indeed we find that the determinant of the matrix
\begin{equation}\label{C1}
{\cal C}=\left(
\begin{array}{cc}
  \{D_\alpha, D_\beta\} & \{D_\alpha, \bar D_{\dot\beta}\} \\
  \{\bar D_{\dot\alpha}, D_\beta\} & \{\bar D_{\dot\alpha}, \bar D_{\dot\beta}\} \\
\end{array}
\right)=
\left(
\begin{array}{cc}
  -2i z\epsilon_{\alpha\beta} & 2ip_{\alpha\dot\beta} \\
  -2ip_{\beta\dot\alpha} & 2i\bar z\epsilon_{\dot\alpha\dot\beta} \\
\end{array}
\right)\,.
\end{equation}
is equal to\footnote{ In calculation it is convenient to use that
$\det{\cal C}=\det D\det (A-BD^{-1}C)$ for matrix ${\cal C}=\left(
\begin{array}{cc}
  A & B \\
  C & D \\
\end{array}
\right)=\left(
\begin{array}{cc}
  1 & B \\
  0 & D \\
\end{array}
\right)
\left(%
\begin{array}{cc}
  A-BD^{-1}C & 0 \\
  D^{-1}C & 1 \\
\end{array}
\right)$.
}
\begin{equation}\label{det}
\det{\cal C}=16 (p^2+|z|^2)^2.
\end{equation}
We see from~(\ref{det}) that the matrix ${\cal C}$
(see~(\ref{C1})) is invertible for any $z$, and the constraints (\ref{D1}--\ref{bD1}) are
 second class.

\subsection{Quantization}
The first quantization of the model can be performed using one of
two methods:

{\bf i)} Following the technique of quantization of systems
 with second class constraints one can introduce Dirac brackets (DB) for the independent
phase space degrees of freedom ${\cal Z}_M=(x^\mu, p_\mu,
\lambda_\alpha, \bar\lambda_{\dot\alpha})$
\begin{equation}\label{DB}
\{{\cal Z}_M,{\cal Z}_N\}^\ast=  \{{\cal Z}_M,{\cal Z}_N\}- \{{\cal
Z}_M,D_r\}({\cal C}^{-1})_{rs}\{D_s,{\cal Z}_N\}
\end{equation}
where $D_r=(D_\alpha, \bar D_{\dot\alpha})$. In particular for
suitably normalized spinor coordinates\footnote{On the mass shell
$T\approx 0$ and at $z=m$ we get $\eta_\alpha =2m\lambda_\alpha$
and $\bar\eta_{\dot\alpha} =2m\bar\lambda_{\dot\alpha}$.}
\begin{equation}\label{sp-var}
\eta_\alpha =[2(p^2+|z|^2)]^{1/2}\lambda_\alpha\,, \qquad
\bar\eta_{\dot\alpha}
=[2(p^2+|z|^2)]^{1/2}\bar\lambda_{\dot\alpha}
\end{equation}
one obtains the relations
\begin{equation}\label{DB-sp}
\{\eta_{\alpha},\eta_{\beta}\}^\ast= -i\bar
z\epsilon_{\alpha\beta} \,,\qquad
\{\bar\eta_{\dot\alpha},\bar\eta_{\dot\beta}\}^\ast=
iz\epsilon_{\dot\alpha\dot\beta} \,,\qquad
\{\eta_{\alpha},\bar\eta_{\dot\beta}\}^\ast= i p_{\alpha\dot\beta}
\end{equation}
leading after quantization to noncommutative Weyl spinor
coordinates. Similarly one can calculate
$$
\{x_{\mu},x_{\nu}\}^\ast={\textstyle\frac{i}{2(p^2+|z|^2)}}S_{\mu\nu}\,,
$$
$$
S^{\mu\nu}=\lambda^\alpha\left[(\sigma^{\mu\nu})_\alpha{}^\beta
p_{\beta\dot\gamma}+
p_{\alpha\dot\beta}(\bar\sigma^{\mu\nu})^{\dot\beta}{}_{\dot\gamma}\right]
\bar\lambda^{\dot\gamma} +z\lambda^\alpha
(\sigma^{\mu\nu})_\alpha{}^\beta \lambda_\beta +\bar z
\bar\lambda_{\dot\alpha}
(\bar\sigma^{\mu\nu})^{\dot\alpha}{}_{\dot\beta}
\bar\lambda^{\dot\beta}\,,
$$
$$
(\sigma^{\mu\nu})_\alpha{}^\beta\equiv {\textstyle\frac{1}{2}}
(\sigma^\mu_{\alpha\dot\gamma}\sigma^{\nu\dot\gamma\beta}-
\sigma^\nu_{\alpha\dot\gamma}\sigma^{\mu\dot\gamma\beta})\,, \qquad
(\bar\sigma^{\mu\nu})^{\dot\alpha}{}_{\dot\beta}\equiv
{\textstyle\frac{1}{2}}( \sigma^{\mu\dot\alpha\gamma}
\sigma^\nu_{\gamma\dot\beta}- \sigma^{\nu\dot\alpha\gamma}
\sigma^\mu_{\gamma\dot\beta})\,,
$$
i.e. we see that the coordinates are becoming also noncommutative.

One can note that after the linear transformation of the form
\begin{equation}
{{\eta}}^\prime_{\alpha}={\eta}_{{\alpha}}+
{c} p_{\alpha\dot\beta} \,\bar\eta^{\dot\beta}
\, ,
\qquad
{\bar{\eta}}^\prime_{\alpha}=\bar{\eta}_{\dot{\alpha}}+
\bar{c}\eta^\beta    p_{\beta\dot\alpha}
\, ,
\end{equation}
we can obtain from the
algebra~(\ref{DB-sp}) for certain choice of $c$ the DB relations
$\{\eta^\prime_{\alpha},\eta^\prime_{\beta}\}^\ast
\sim\epsilon_{\alpha\beta}$,
$\{\eta^\prime_{\alpha},\bar\eta^\prime_{\dot\beta}\}^\ast=0$. The
algebra of such type is used for description of massless fields
with arbitrary helicities in [3,4]. For other choice of $c$ we
obtain alternatively
$\{\eta^\prime_{\alpha},\eta^\prime_{\beta}\}^\ast=0$,
$\{\eta^\prime_{\alpha},\bar\eta^\prime_{\dot\beta}\}^\ast \sim
p_{\alpha\dot\beta}$. In such a case $\eta^\prime_{\alpha}$ and
$\bar\eta^\prime_{\dot\alpha}$ can be treated as  of suitably rescaled
creation and annihilation operators.

{\bf ii)} Other way is the Gupta--Bleuler quantization method. Such a
technique implies the split of the second class constraints into
complex--conjugated pairs, with holomorphic and antiholomorphic
parts forming separately the subalgebras of first class constraints.
The algebra~(\ref{al1}--\ref{al2}) of the constraints~(\ref{D1}), (\ref{bD1})
does not satisfy these requirements. Let us introduce, however,
new constraints as follows
\begin{equation}\label{calD}
{\cal D}_\alpha=D_\alpha+ {\textstyle\frac{b}{\bar z
}}p_{\alpha\dot\beta}\bar D^{\dot\beta}\,, \qquad \bar{\cal
D}_{\dot\alpha}=\bar D_{\dot\alpha}+ {\textstyle\frac{\bar
b}{z}}D^{\beta}p_{\beta\dot\alpha}\,,
\end{equation}
$\bar{\cal D}_{\dot\alpha}=(\overline{{\cal D}_\alpha})$. If $b$
satisfies the equation $(b^2-2b)\frac{m^2}{|z|^2}-1=0$ (i.e.
 $b=
 (1\pm\sqrt{1+\frac{|z|^2}{m^2}})$) the
algebra of the constraints~(\ref{calD}) takes the form
$$
\{{\cal D}_\alpha, {\cal D}_\beta \}= {\textstyle\frac{2i}{\bar z
}}\epsilon_{\alpha\beta}T\,,\qquad \{\bar{\cal D}_{\dot\alpha},
\bar{\cal D}_{\dot\beta}\}=-{\textstyle\frac{2i}{z}}
\epsilon_{\dot\alpha\dot\beta}T \,,\qquad \{{\cal D}_\alpha, \bar
{\cal D}_{\dot\beta}\}=-4b(1+{\textstyle\frac{m^2}{|z|^2}})
ip_{\alpha\dot\beta}
-{\textstyle\frac{2b^2i}{|z|^2}}p_{\alpha\dot\beta}T\,.
$$
We see that the constraints~(\ref{calD}) are suitable for application
of Gupta--Bleuler quantization method. It should be mentioned that the
transformation from constraints $(D_\alpha,\bar D_{\dot\alpha})$ to
constraints $({\cal D}_\alpha, \bar{\cal D}_{\dot\alpha})$ is
invertible.

We shall assume that the wave function satisfies the Klein--Gordon
equation, what follows from the constraint~(\ref{T1}). On the mass
shell ~(\ref{T1}) the constraints~(\ref{calD}) have the form
\begin{equation}\label{calD1}
{\cal D}_\alpha = \pi^\prime_\alpha
-2b(1+{\textstyle\frac{m^2}{|z|^2}})ip_{\alpha\dot\beta}\bar\lambda^{\prime\dot\beta}\approx
0\,,\qquad \bar{\cal D}_{\dot\alpha}= \bar\pi^\prime_{\dot\alpha}
+2b(1+{\textstyle\frac{m^2}{|z|^2}})i\lambda^{\prime\beta}
p_{\beta\dot\alpha}\approx 0
\end{equation}
where we introduced new spinor variables via the following canonical
transformation
\begin{equation}\label{pr-p}
\pi^\prime_\alpha\equiv \pi_\alpha+ {\textstyle\frac{b}{\bar
z}}p_{\alpha\dot\beta}\bar\pi^{\dot\beta} \,,\qquad
\bar\pi^\prime_{\dot\alpha}\equiv \bar \pi_{\dot\alpha}+
{\textstyle\frac{b}{z}}\pi^{\beta}p_{\beta\dot\alpha}\,,
\end{equation}
\begin{equation}\label{pr-l}
\lambda^{\prime\alpha}\equiv
{\textstyle\frac{|z|^2}{|z|^2+b^2p^2}}(\lambda^\alpha-
{\textstyle\frac{b}{z}}\bar\lambda_{\dot\beta}p^{\dot\beta\alpha})\,,\qquad
\bar\lambda^{\prime\dot\alpha}\equiv
{\textstyle\frac{|z|^2}{|z|^2+b^2p^2}}(\bar\lambda^{\dot\alpha}-
{\textstyle\frac{b}{\bar z}}p^{\dot\alpha\beta}\lambda_{\beta})
\end{equation}
i.e. we obtain the standard canonical commutation relations
(compare with~(\ref{CCR}))
\begin{equation}\label{CCR-pr}
\{\lambda^{\prime\alpha},
\pi^\prime_\beta\}=\delta^\alpha_\beta\,,\qquad
\{{\bar\lambda}^{\prime\dot\alpha},\bar\pi^\prime_{\dot\beta}\}=
\delta^{\dot\alpha}_{\dot\beta}\,,\qquad \{\lambda^{\prime\alpha},
\bar\pi^\prime_{\dot\beta}\}=\{{\bar\lambda}^{\prime\dot\alpha},
\pi^\prime_\beta\}=0\,.
\end{equation}

For the  quantization of our model we consider the Schr\"{o}dinger
representation of the CCR~(\ref{CCR-pr})
\begin{equation}\label{rep}
\pi^\prime_{\alpha}=-i{\partial}/{\partial\lambda^{\prime\alpha}}\,,
\qquad \bar\pi^\prime_{\dot\alpha}=
-i{\partial}/{\partial\bar\lambda^{\prime\dot\alpha}}
\end{equation}
and use the wave function $\Psi$ in the momentum representation, i.e.
  $\Psi=\Psi(p_\mu,\lambda^{\prime\alpha},
\bar\lambda^{\prime\dot\alpha})$. The spinorial wave equation
$\bar{\cal D}_{\dot\alpha}\Psi=0$ takes the following
form\footnote{The choice of ${\cal D}_{\alpha}$ in place $\bar{\cal
D}_{\dot\alpha}$ is equally well possible.}
\begin{equation}\label{equa}
(-{\partial}/{\partial\bar\lambda^{\prime\dot\alpha}}+
2b(1+{\textstyle\frac{m^2}{|z|^2}})\lambda^{\prime\beta}
p_{\beta\dot\alpha})\Psi=0\,.
\end{equation}
The solution of~(\ref{equa}) is given by
\begin{equation}\label{sol}
\Psi(p_\mu,\lambda^{\prime\alpha},
\bar\lambda^{\prime\dot\alpha})=e^{2b
(1+{\textstyle\frac{m^2}{|z|^2}})\lambda^{\prime\beta}
p_{\beta\dot\alpha}\bar\lambda^{\prime\dot\alpha}}
\tilde\Psi(p_\mu,\lambda^{\prime\alpha})
\end{equation}
where the field $\tilde\Psi(p_\mu,\lambda^{\prime\alpha})$ depends only
on one Weyl spinor $\lambda^{\prime\alpha}$ and provides the bosonic
counterpart of $D=4$ $N=1$ chiral superfield.

Due to the bosonic nature of $\lambda^{\prime\alpha}$ in expansion of
$\tilde\Psi(p_\mu,\lambda^{\prime\alpha})$ there is an infinite number
of space--time fields
$\psi_{\alpha_1\cdots\alpha_n}(p)=\psi_{(\alpha_1\cdots\alpha_n)}(p)$,
$n=0,1,\dots,\infty$. The mass--shell condition~(\ref{T1}) after the
transition by Fourier transformation to the space--time picture, leads
to the Klein--Gordon (KG) equation ($\Box\equiv\partial_\mu\partial^\mu$)
\begin{equation}\label{sol1}
(\Box +m^2)\Psi(x;\lambda,\bar\lambda)=0 \qquad\Leftrightarrow\qquad
(\Box +m^2)\psi_{\alpha_1\cdots\alpha_n}(x)=0\quad (n=0,1,2,\dots)\,.
\end{equation}
Here we should observe that

i) The half-integer spin fields (n odd) satisfy KG equation, however in massless
 case the half-integer helicity fields do satisfy linear equations (see Sect. 3),

 ii) The spin--statistic theorem
  is not valid
--
both integer and half--integer spin fields are bosonic. We shall come back
to the question of statistics in Sect. 4.

\section{Massless particle model with $N=1$ bosonic counterpart of SUSY}

The model~(\ref{act}), (\ref{Lagr1}) can be described equivalently
by the Lagrangian
\begin{equation}\label{Lagr1a}
{\cal L}=-{\textstyle\frac{1}{2e}}(\dot\omega_\mu\dot\omega^\mu+
e^2 m^2) -i (z\dot\lambda^\alpha\lambda_\alpha - \bar
z\bar\lambda_{\dot\alpha}\dot{\bar\lambda}{}^{\dot\alpha})\,.
\end{equation}
After eliminating the einbein $e$ by means of its equation of motion,
from ~(\ref{Lagr1a}) one obtains the Lagrangian~(\ref{Lagr1}).
 The massless limit of (\ref{Lagr1a}) looks as follows
\begin{equation}\label{Lagrm=0}
{\cal L}=-{\textstyle\frac{1}{2e}}\dot\omega_\mu\dot\omega^\mu- i
(z\dot\lambda^\alpha\lambda_\alpha - \bar
z\bar\lambda_{\dot\alpha}\dot{\bar\lambda}{}^{\dot\alpha})\,.
\end{equation}
Besides the constraint $p_e \approx 0$ which implies pure gauge
character of the einbein $e$, from (\ref{Lagrm=0}) one gets the
 following constraints
\begin{equation}\label{Tm=0}
T = p^2 \approx 0\,,
\end{equation}
\begin{equation}\label{Dm=0}
D_\alpha = \pi_\alpha
+ip_{\alpha\dot\beta}\bar\lambda^{\dot\beta}+i
z\lambda_\alpha\approx 0\,,\qquad \bar D_{\dot\alpha}=
\bar\pi_{\dot\alpha} - i\lambda^\beta p_{\beta\dot\alpha} -i\bar
z\bar\lambda_{\dot\alpha}\approx 0\,.
\end{equation}
The nonvanishing Poisson brackets are
\begin{equation}\label{al10}
\{D_\alpha, \bar D_{\dot\beta}\}=2ip_{\alpha\dot\beta}\,, \qquad
\{D_{\alpha}, D_{\beta}\}=-2i z\epsilon_{\alpha\beta}\,,\qquad
\{\bar D_{\dot\alpha}, \bar D_{\dot\beta}\}= 2i \bar z
\epsilon_{\dot\alpha\dot\beta}\,.
\end{equation}
The mass constraint~(\ref{Tm=0}) is of the first class. The determinant of the
 Poisson brackets
matrix~(\ref{C1})  characterizing  the spinorial
constraints~(\ref{Dm=0}) is the following
\begin{equation}\label{det0}
\det{\cal C}=16 (p^2+|z|^2)^2 \approx 16 |z|\,^4.
\end{equation}
If $z\neq 0$ all spinorial constraints~(\ref{Dm=0}) are
second class. In the case of vanishing central charges $z= 0$ the four
spinorial constraints~(\ref{Dm=0}) contain two second class
constraints and two first class. Below  we analyze
massless particle at $z= 0$ with spinorial first class
constraints, defined as follows:

\begin{equation}\label{first}
F^{\dot\alpha}=p^{\dot\alpha\beta}D_\beta \approx 0\,,\qquad \bar
F^{\alpha}= \bar D_{\dot\beta}p^{\dot\beta\alpha} \approx 0
\end{equation}
with the following Poisson brackets
$$
\{F^{\dot\alpha}, \bar
D_{\dot\beta}\}=2i\delta^{\dot\alpha}_{\dot\beta} T\approx 0\,,
\qquad \{\bar F^{\alpha}, D_{\beta}\}= -2i\delta^{\alpha}_{\beta}
T\approx 0
$$
 But the first class constraints~(\ref{first}) are
reducible: $p_{\alpha\dot\beta}F^{\dot\beta}\approx 0$, $\bar
F^{\beta}p_{\beta\dot\alpha}\approx 0$. The irreducible separation of
first and second class constraints is obtained by projecting of
the spinorial constraints~(\ref{Dm=0}) along spinors $\lambda^\alpha$
and $\bar\lambda_{\dot\alpha}p^{\dot\alpha\alpha}$.\footnote{This
procedure is corrected since spinors $\lambda^\alpha$ and
$\bar\lambda_{\dot\alpha}p^{\dot\alpha\alpha}$ are not
proportional in considered task. Otherwise, when
$\lambda^{\alpha}p_{\alpha\dot\alpha}\bar\lambda^{\dot\alpha}=0$,
we have $p_{\alpha\dot\alpha}
\sim\lambda_{\alpha}\bar\lambda_{\dot\alpha}$. Then the spinorial
constraints~(\ref{Dm=0}), taking the form $\pi_\alpha \approx 0$,
$\bar\pi_{\dot\alpha} \approx 0$, exclude completely the
dependence on $\lambda$, $\bar\lambda$. As result we obtain the
system describing only by the variables $x^\mu$, $p_\mu$ and the
constraint~(\ref{Tm=0}) i. e. the massless particle of zero
helicity.} The constraints
\begin{equation}\label{cons-G}
G\equiv\lambda^{\alpha} D_\alpha \approx 0\,,\qquad\qquad \bar G
\equiv\bar D_{\dot\alpha}\bar\lambda^{\dot\alpha} \approx 0
\end{equation}
are  second class whereas the constraints
\begin{equation}\label{cons-F}
F\equiv\bar\lambda_{\dot\alpha}p^{\dot\alpha\alpha} D_\alpha
\approx 0\,,\qquad\qquad \bar F \equiv\bar
D_{\dot\alpha}p^{\dot\alpha\alpha}\lambda_{\alpha} \approx 0
\end{equation}
are of first class. Their Poisson brackets look as follows:
$$
\{G, \bar G \}=2i\lambda^{\alpha}
p_{\alpha\dot\alpha}\bar\lambda^{\dot\alpha}\not\approx 0\,,\qquad
\{F, \bar F \}=-(\lambda^{\alpha}\pi_{\alpha}-
\bar\pi_{\dot\alpha}\bar\lambda^{\dot\alpha})\,T\,,
$$
$$
\{G, F \}= -\{\bar G, F \}=F\,,\qquad \{G, \bar F \}=  -\{\bar G,
\bar F \}=-\bar F \,.
$$

We carry out quantization of massless particle with $N=1$ bosonic
counterpart of SUSY by Gupta--Bleuler method. The wave equations are
imposed by the first class constraints~(\ref{Tm=0}), (\ref{cons-F})
$T\approx 0$, $F\approx 0$, $\bar F\approx 0$ and
 either $\bar G\approx 0$
or $G\approx 0$. But the pair of constraints $G\approx 0$ and $F\approx 0$
are equivalent to the constraints $D_{\alpha}\approx 0$;
 similarly the constraints $\bar G\approx
0$ and $\bar F\approx 0$ are equivalent to the constraints $\bar
D_{\dot\alpha}\approx 0$. Thus we have two possible quantizations:
\begin{description}
    \item -- `bosonic chiral'  
     quantization with the wave equations
    \begin{equation}\label{chir0}
    T|\Psi\rangle=0\,, \qquad F|\Psi\rangle=0\,, \qquad
    \bar D_{\dot\alpha}|\Psi\rangle=0
    \end{equation}
    \item -- `bosonic antichiral'  quantization with  wave function subjected
     to the conditions
    \begin{equation}\label{achir0}
    T|\Psi\rangle=0\,, \qquad \bar F|\Psi\rangle=0\,,
    \qquad D_{\alpha}|\Psi\rangle=0\,.
    \end{equation}
\end{description}

Let us consider the chiral case. In the representation
$$
p_\mu=-i{\partial}/{\partial x^{\mu}}\equiv-i{\partial}_\mu\,,
\qquad
\pi_{\alpha}=-i{\partial}/{\partial\lambda^{\alpha}}\equiv-i{\partial}_\alpha\,,
\qquad \bar\pi_{\dot\alpha}=
-i{\partial}/{\partial\bar\lambda^{\dot\alpha}}\equiv-i\bar{\partial}_{\dot\alpha}
$$
the wave function $\Psi(x,\lambda,\bar\lambda)$ satisfies the
equations
\begin{equation}\label{mass0}
\Box\,\Psi=0\,,
\end{equation}
\begin{equation}\label{chir-0}
\bar D_{\dot\alpha}\,\Psi= (-i\bar\partial_{\dot\alpha} -
\lambda^\beta \partial_{\beta\dot\alpha})\,\Psi=0
\end{equation}
\begin{equation}\label{chir-0-0}
-i\bar\lambda_{\dot\alpha}\partial^{\dot\alpha\alpha}
D_\alpha\,\Psi=
-\bar\lambda_{\dot\alpha}\partial^{\dot\alpha\alpha}
\partial_\alpha\,\Psi=0
\end{equation}
In the variables $x^\mu_L=x^\mu+i\lambda\sigma^\mu\bar\lambda$,
$\lambda^\alpha$, $\bar\lambda^{\dot\alpha}$  bosonic SUSY-covariant
derivatives take the form
\begin{equation}\label{D-L}
D_\alpha= -i\partial_\alpha
+2\partial_{L\alpha\dot\alpha}\bar\lambda^{\dot\alpha}\,,\qquad\qquad
\bar D_{\dot\alpha}= -i\bar\partial_{\dot\alpha} \,.
\end{equation}
Thus due to the chirality condition~(\ref{chir-0}) the wave
function does not depend on $\bar\lambda^{\dot\alpha}$. It
 depends only on the left chiral variables
$z_L^{}=(x^\mu_L,\lambda^\alpha)$,
 and commuting spinor $\lambda$.
 One can write the following expansion
\begin{equation}\label{wf-exp}
\Psi(x^{}_L,\lambda)=\sum_{n=0}^{\infty}
\lambda^{\alpha_1}\ldots\lambda^{\alpha_n}\phi_{\alpha_1\ldots\alpha_n}(x^{}_L)
\end{equation}
where the multispinor fields are totally symmetric in spinor
indices, i.e.
$\phi_{\alpha_1\ldots\alpha_n}=\phi_{(\alpha_1\ldots\alpha_n)}$.
The usual fields depending on real space--time coordinates $x^\mu$ are
obtained by
$$
\phi_{\alpha_1\ldots\alpha_n}(x)=e^{-i\lambda\sigma^\mu\bar\lambda\partial_\mu}
\phi_{\alpha_1\ldots\alpha_n}(x^{}_L)\,.
$$
The equation~(\ref{chir-0-0}) gives Fierz--Pauli equations for the
component fields
\begin{equation}\label{PF}
\partial^{\dot\beta\beta}\,\phi_{\beta\alpha_2\ldots\alpha_n}=0\,.
\end{equation}
The Klein--Gordon equation $\Box\,\phi_{\alpha_1\ldots\alpha_n} =0$,
resulting from~(\ref{mass0}),  follows also from~(\ref{PF}).
We see therefore that the expansion of the
wave function~(\ref{wf-exp})  describes an infinite set of massless particles
with helicities $n/2$.

The Gupta-Bleuler quantization procedure presented here is analogous to the one
 used for the quantization of massless Brink-Schwarz  superparticle, but due to the bosonic
 character of spinorial variable $\lambda_\alpha$ we get infinite helicity spectrum.
We recall that
the infinite set of integer and half--integer helicities
 describes also the
 spectrum of supersymmetric massless particles propagating in
 tensorial superspace~\cite{BandLuk}.

\section{Massive relativistic particles with $N=2$ bosonic counterpart of SUSY.}

\subsection{$N=2$ action and the constraints}
Let us introduce two commuting Weyl spinors $\lambda^\alpha_i$,
$\bar\lambda^{\dot\alpha}_i= (\overline{\lambda^\alpha_i})$ ($i=1,2$). The
natural generalization of the Lagrangian~(\ref{Lagr1}) is
\begin{equation}\label{Lagr2}
{\cal L}=-m(\dot\omega_\mu\dot\omega^\mu)^{1/2} -i
(z_{ij}\dot\lambda^\alpha_i\lambda_{\alpha j} - \bar
z_{ij}\bar\lambda_{\dot\alpha
j}\dot{\bar\lambda}{}^{\dot\alpha}_i)\,.
\end{equation}
Here the constant matrix $z_{ij}$ is symmetric, $z_{ij}=z_{ji}$;
 the last terms in~(\ref{Lagr2}) are total derivatives, e.g.
$z_{ij}\dot\lambda^\alpha_i\lambda_{\alpha j}=\frac{1}{2}
(z_{ij}\lambda^\alpha_i\lambda_{\alpha j})$  if
$z_{ij}=-z_{ji}$.

The $\omega$--form can be written in general case as follow
\begin{equation}\label{om2}
\dot\omega^\mu=\dot x^\mu- i\kappa_{ij}
(\dot\lambda^\alpha_i\sigma^\mu_{\alpha\dot\beta}\bar\lambda^{\dot\beta}_j-
 \lambda^\alpha_j\sigma^\mu_{\alpha\dot\beta}\dot{\bar\lambda}^{\dot\beta}_i)
\end{equation}
where $\kappa_{ij}=\kappa_{ji}$ is the $2\times 2$ Hermitean metric
 in $N=2$ unitary space.
 If we consider possible linear definitions of spinors
$\lambda^\alpha_i$ in $N=2$ internal space one can choose
\begin{equation}\label{eqfl37}
\kappa_{ij}=\left(
\begin{array}{cc}
  1 & 0 \\
  0 & \kappa \\
\end{array}
\right)
\end{equation}
 where $\kappa$ is real.

From expressions for the canonical momenta
\begin{equation}\label{px2}
p_\mu=\frac{\partial {\cal L}}{\partial\dot x^\mu}=
-m(\dot\omega_\nu\dot\omega^\nu)^{-1/2}\dot\omega_\mu\,,
\end{equation}
\begin{equation}\label{pi2}
\pi_{\alpha i} =\frac{\partial {\cal
L}}{\partial\dot\lambda^\alpha_i}=
-i\kappa_{ij}p_{\alpha\dot\beta}\bar\lambda^{\dot\beta}_j-iz_{ij}\lambda_{\alpha
j}\,,
\end{equation}
\begin{equation}\label{bpi2}
\bar\pi_{\dot\alpha i} =\frac{\partial {\cal
L}}{\partial\dot{\bar\lambda}{}^{\dot\alpha}_i}=
i\kappa_{ij}\lambda^\beta_j p_{\beta\dot\alpha} +i\bar z_{ij}
\bar\lambda_{\dot\alpha j}
\end{equation}
we obtain the following constraints
\begin{equation}\label{T2}
T \equiv p^2 -m^2\approx 0\,,
\end{equation}
\begin{equation}\label{D2}
D_{\alpha i} \equiv \pi_{\alpha i}
+i\kappa_{ij}p_{\alpha\dot\beta}\bar\lambda^{\dot\beta}_j+iz_{ij}\lambda_{\alpha
j}\approx 0\,,
\end{equation}
\begin{equation}\label{bD2}
\bar D_{\dot\alpha i}\equiv \bar\pi_{\dot\alpha i} -
i\kappa_{ij}\lambda^\beta_j p_{\beta\dot\alpha} -i\bar z_{ij}
\bar\lambda_{\dot\alpha j}\approx 0\,.
\end{equation}

Using the canonical Poisson brackets
$$
\{x^\mu, p_\nu\}=\delta^\mu_\nu\,,\qquad \{\lambda^\alpha_i,
\pi_{\beta j}\}=\delta^\alpha_\beta\delta_{ij}\,,\qquad
\{{\bar\lambda}^{\dot\alpha}_i,\bar\pi_{\dot\beta j}\}=
\delta^{\dot\alpha}_{\dot\beta}\delta_{ij}
$$
($\{\lambda_{\alpha i},  \pi_{\beta j}\}=\{\pi_{\alpha i},
\lambda_{\beta j}\}= -\epsilon_{\alpha\beta}\delta_{ij}$,
$\{{\bar\lambda}_{\dot\alpha i},\bar\pi_{\dot\beta j}\}=
\{{\bar\pi}_{\dot\alpha i},\bar\lambda_{\dot\beta j}\}=
-\epsilon_{\dot\alpha\dot\beta}\delta_{ij}$) we obtain nonzero
Poisson brackets of the constraints~(\ref{T2})-(\ref{bD2})
\begin{equation}\label{al1-2}
\{D_{\alpha i}, \bar D_{\dot\beta
j}\}=2i\kappa_{ij}p_{\alpha\dot\beta}\,,
\end{equation}
\begin{equation}\label{al2-2}
\{D_{\alpha i}, D_{\beta j}\}=-2i
z_{ij}\epsilon_{\alpha\beta}\,,\qquad \{\bar D_{\dot\alpha i},
\bar D_{\dot\beta j}\}= 2i\bar z_{ij}
\epsilon_{\dot\alpha\dot\beta}\,.
\end{equation}
It should be pointed out that the relations~(\ref{al1-2}),
(\ref{al2-2}) with changed sign on the rhs describe the bosonic
counterpart of the generalized $N=2$ superalgebra with the Hermitean  metric
$\kappa_{ij}$ in internal $N=2$ space.

The constraint~(\ref{T2}) $T\approx 0$ is the first class constraint.
 From the spinor constraints~(\ref{D2}),
(\ref{bD2})
 one gets  the following $4 \times 4$ matrix of PB
\begin{equation}\label{C2}
{\cal C}=\left(
\begin{array}{cc}
  \{D_{\alpha i}, D_{\beta j}\} & \{D_{\alpha i}, \bar D_{\dot\beta j}\} \\
  \{\bar D_{\dot\alpha i}, D_{\beta j}\} & \{\bar D_{\dot\alpha i}, \bar D_{\dot\beta j}\} \\
\end{array}
\right)= \left(
\begin{array}{cc}
  -2i z_{ij}\epsilon_{\alpha\beta} & 2i\kappa_{ij}p_{\alpha\dot\beta} \\
  -2i\kappa_{ij}p_{\beta\dot\alpha} & 2i\bar z_{ij} \epsilon_{\dot\alpha\dot\beta} \\
\end{array}
\right)\,.
\end{equation}
We obtain that
$$
\det{\cal C}=2^8[\det(\hat z\hat{\bar z}+p^2\hat\kappa\hat{\bar
z}^{-1}\hat\kappa\hat{\bar z})]^2
$$
where `hats' denote the corresponding matrices, i.e. $\hat z=(z_{ij})$,
$\hat{\bar z}=(\bar z_{ij})$ and $\hat\kappa=(\kappa_{ij})$ is given by (\ref{eqfl37}).
One can consider two cases:

{\sl i)} If matrix $\hat z=(z_{ij})$ is diagonal, $(z_{ij})=\left(
\begin{array}{cc}
  z_1 & 0 \\
  0 & z_2 \\
\end{array}
\right)$, we obtain that $\det(\hat z\hat{\bar
z}+p^2\hat\kappa\hat{\bar z}^{-1}\hat\kappa\hat{\bar
z})=(|z_1|^2+p^2)(|z_2|^2+p^2\kappa^2)$,  i.e. it is always nonvanishing.
 We see therefore that for  arbitrary
 values of $\kappa$ and $z_1, z_2$  all
  the constraints~(\ref{D2}), (\ref{bD2}) are  second class.

{\sl ii)} In case of antidiagonal matrix $(z_{ij})=\left(
\begin{array}{cc}
  0 & z \\
  z & 0 \\
\end{array}
\right)$ (we remind that matrix $z_{ij}$ is symmetric), we
obtain that $\det(\hat z\hat{\bar z}+p^2\hat\kappa\hat{\bar
z}^{-1}\hat\kappa\hat{\bar z})=(|z|^2+p^2\kappa)^2$.
One gets that the matrix of Poisson brackets of the constraints~(\ref{C2})
has vanishing determinant if
$\kappa=-\frac{|z|^2}{m^2}<0$
 and we conclude that in such a case the
first class constraints are present in the
model.
Putting $z=m$, i.e. $\kappa=-1$, it is easy to  check that the unitary metric tensor
 $\kappa_{ij}$ implies the invariance of the form $\omega_\mu$
 (see (\ref{om2})) under $U(1,1)$ symmetry. The presence of the central charge reduces
  however this symmetry to  the invariance group $O(1,1) = U(1,1) \cap O(2;c)$,
   and only in this case the first class constraints are present in the model
    (\ref{Lagr2}).\footnote{We recall that in case of standard
$N=2$ superparticle when spinor variables are Grassmannian and the
matrix $z_{ij}$ is skew--symmetric, $(z_{ij})=\left(
\begin{array}{cc}
  0 & z \\
  -z & 0 \\
\end{array}
\right)$, the first class constraints are presented (the matrix of
Poisson brackets of the constraints has vanishing determinant) if
$\kappa=\frac{|z|^2}{m^2}>0$ and the internal $N=2$ symmetry in the presence
 of central charges $z=m$ is $U(2)\cap Sp(2;c) = SU(2)$.}

 In  case {\sl ii)} we will  consider a simple choice $z=m$, i.e.
$\kappa=-\frac{|z|^2}{m^2}=-1$. Introducing the notations
$\lambda^\alpha_1\equiv\lambda^\alpha$ and
$\lambda^\alpha_2\equiv\eta^\alpha$ the Lagrangian~(\ref{Lagr2})
and $\omega$--form~(\ref{om2}) are
\begin{equation}\label{Lagr2a}
{\cal L}=-m(\dot\omega_\mu\dot\omega^\mu)^{1/2} -i
m(\dot\lambda^\alpha\eta_{\alpha}+\dot\eta^\alpha\lambda_{\alpha} -
\bar\lambda_{\dot\alpha}\dot{\bar\eta}{}^{\dot\alpha}
-\bar\eta_{\dot\alpha}\dot{\bar\lambda}{}^{\dot\alpha})\,,
\end{equation}
\begin{equation}\label{om2a}
\dot\omega^\mu=\dot x^\mu- i
(\dot\lambda^\alpha\sigma^\mu_{\alpha\dot\beta}\bar\lambda^{\dot\beta}-
 \lambda^\alpha\sigma^\mu_{\alpha\dot\beta}\dot{\bar\lambda}^{\dot\beta})
+i (\dot\eta^\alpha\sigma^\mu_{\alpha\dot\beta}\bar\eta^{\dot\beta}-
 \eta^\alpha\sigma^\mu_{\alpha\dot\beta}\dot{\bar\eta}^{\dot\beta})\,.
\end{equation}

\subsection{Description of the model in terms of Dirac spinors}

The formulation~(\ref{Lagr2a}) has an attractive interpretation if
we pass to the commuting four--component  Dirac spinor
$$
\psi_a= \left(
         \begin{array}{c}
           \lambda_\alpha \\
           {\bar\eta}^{\dot\alpha} \\
         \end{array}
       \right)
$$
where $a=1,2,3,4$. The  Dirac matrices
$(\gamma_\mu)_a{}^b$ in Weyl representation are as follows
$$
(\gamma_\mu)_a{}^b=\left(
             \begin{array}{cc}
               0 & \sigma^\mu_{\alpha\dot\beta} \\
               \sigma^{\mu{\dot\alpha\beta}} & 0 \\
             \end{array}
           \right)\,, \qquad \{\gamma_\mu, \gamma_\nu\}=2\eta_{\mu\nu}
$$
where
$\sigma^0_{\alpha\dot\beta}=\sigma^{0{\dot\alpha\beta}}=1_2$ and
$\sigma^i_{\alpha\dot\beta}=-\sigma^{i{\dot\alpha\beta}}$
 $(i=1,2,3)$  are the Pauli matrices. Then
$$
\bar\psi^a=(\psi^+\gamma_0)^a=(\eta^\alpha,
\bar\lambda_{\dot\alpha})
$$
and we obtain
$$
\dot{\bar\psi}\psi-{\bar\psi}\dot\psi=\dot\lambda^\alpha\eta_{\alpha}+
\dot\eta^\alpha\lambda_{\alpha} -
\bar\lambda_{\dot\alpha}\dot{\bar\eta}{}^{\dot\alpha}
-\bar\eta_{\dot\alpha}\dot{\bar\lambda}{}^{\dot\alpha}\,,
$$
$$
\dot{\bar\psi}\gamma^\mu\psi-{\bar\psi}\gamma^\mu\dot\psi=
\dot\eta^\alpha\sigma^\mu_{\alpha\dot\beta}\bar\eta^{\dot\beta} -
\dot\lambda^\alpha\sigma^\mu_{\alpha\dot\beta}\bar\lambda^{\dot\beta}-
(\eta^\alpha\sigma^\mu_{\alpha\dot\beta}\dot{\bar\eta}^{\dot\beta}-
\lambda^\alpha\sigma^\mu_{\alpha\dot\beta}\dot{\bar\lambda}^{\dot\beta})\,.
$$
Thus the Lagrangian~(\ref{Lagr2a}) takes in the notation using Dirac
spinor $\psi$ the following simple form
\begin{equation}\label{Lagr2D}
{\cal L}=-m(\dot\omega_\mu\dot\omega^\mu)^{1/2} -i
m(\dot{\bar\psi}\psi-{\bar\psi}\dot\psi)\,,
\end{equation}
where
\begin{equation}\label{om2D}
\dot\omega^\mu=\dot x^\mu+ i
(\dot{\bar\psi}\gamma^\mu\psi-{\bar\psi}\gamma^\mu\dot\psi)\, .
\end{equation}
We would like to point out that the model with spinorial
variables described by Dirac spinor corresponds to the choice of
noncompact internal sector, with the metric $\kappa_{ij}={\rm
diag}(1,-1)$. It should be added that the model~(\ref{Lagr2D}) in
different context has been firstly proposed in~\cite{ZimFed}.

\subsection{Gupta-Bleuler quantization of the model}

The constraints~(\ref{T2})--(\ref{bD2})
 for $z = m$ or equivalently $\kappa=-1$,
 written
in Dirac notation, are the following
\begin{equation}\label{T-D}
T \equiv p^2 -m^2\approx 0\,,
\end{equation}
\begin{equation}\label{D-D}
D^a \equiv \pi^a +i\bar\psi^b(\hat p-m)_b{}^a\approx 0\,,
\end{equation}
\begin{equation}\label{bD-D}
\bar D_a\equiv \bar\pi_a - i(\hat p-m)_a{}^b \psi_b \approx 0\,.
\end{equation}
Here $\pi^a$ and $\bar\pi_a$ defined as $\pi^{a}=\partial{\cal
L}/\partial\dot{\psi}_a$ and $\bar\pi_{a}=\partial{\cal
L}/\partial\dot{\bar\psi}^a$ are conjugate momenta of $\psi_a$
and ${\bar\psi}^a$; their Poisson brackets are $\{\psi_a,
\pi^b\}=\delta_a^b$ and $\{{\bar\psi}^a, \bar\pi_b\}=\delta^a_b$.
Also we shall use notation $\hat p\equiv \gamma^\mu p_\mu$.

From Poisson brackets of the constraints
\begin{equation}\label{PB-2D}
\{\bar D_a, D^b \}=-2i(\hat p-m)_a{}^b\,, \qquad \{D^a, D^b
\}=0\,,\qquad \{\bar D_a, \bar D_b \}=0\,,
\end{equation}
\begin{equation}\label{PB-2Da}
\{T, D_a \}=\{T,\bar D_a\}=0
\end{equation}
we obtain directly that the constraint~(\ref{T-D}) and the half of
the spinorial constraints
 ~(\ref{D-D}), (\ref{bD-D}) are first class
constraints.

The separation of first and second class  spinorial
constraints in (\ref{D-D}), (\ref{bD-D}) is achieved by the projectors
${\cal P}_\pm\equiv \frac{1}{2m}(m\pm\hat p)$ where $1=({\cal P}_+
+{\cal P}_-)$. One can check that on mass shell $p^2=m^2$ we obtain ${\cal
P}_\pm{\cal P}_\pm={\cal P}_\pm$, ${\cal P}_+{\cal P}_-=0$. From
eight real spinorial constraints~(\ref{D-D}), (\ref{bD-D}) we
construct the following sets of reducible constraints
\begin{equation}\label{F}
F^a =D^b(\hat p+m)_b{}^a\,, \qquad \bar F_a =(\hat p+m)_a{}^b \bar
D_b\,;
\end{equation}
\begin{equation}\label{G}
G^a =D^b(\hat p-m)_b{}^a\,, \qquad \bar G_a =(\hat p-m)_a{}^b \bar
D_b\,.
\end{equation}
Due to the relations
$$
F^b(\hat p-m)_b{}^a=0\,, \qquad (\hat p-m)_a{}^b \bar F_b=0\,;
$$
$$
G^b(\hat p+m)_b{}^a=0\,, \qquad (\hat p+m)_a{}^b \bar D_b=0
$$
on the mass--shell
 (\ref{T-D}) in the set of the constraints
$(F^a, \bar F_a)$ there are only four real independent constraints.
Analogously, the constraints $(G^a, \bar G_a)$ contain as well four
real independent constraints. Expressing  the
constraints~(\ref{D-D}), (\ref{bD-D}) in term of the
constraints~(\ref{F}), (\ref{G}) we get
$$
D^a={\textstyle\frac{1}{2m}}(F^a-G^a)\,, \qquad \bar
D_a={\textstyle\frac{1}{2m}}(\bar F_a-\bar G_a)\,.
$$

The constraints~(\ref{F}), (\ref{G}) satisfy the following Poisson
brackets algebra
$$
\{\bar F_a, F^b \}=-2i(\hat p+m)_a{}^b T\,,\qquad \{F^a, F^b
\}=\{\bar F_a, \bar F_b \}=0\,,
$$
$$
\{\bar F_a, G^b \}=\{\bar G_a, F^b \}=-2i(\hat p-m)_a{}^b
T\,,\qquad\{\bar F_a, \bar G_b \}=\{ F^a, G^b \}=0\,,
$$
$$
\{\bar G_a, G^b \}=-8im^2(\hat p+m)_a{}^b -2i[2m\delta_a^b+ (\hat
p+m)_a{}^b]T\,,\qquad\{ G^a, G^b \}=\{\bar G_a, \bar G_b \}=0\,.
$$
From eight real spinorial constraints present in~(\ref{D-D}), (\ref{bD-D})
four  independent constraints in $(F^a, \bar F_a)$ are first
class whereas  four independent constraints contained in  $(G^a, \bar
G_a)$ are second class.

 We shall employ the  Gupta--Bleuler quantization method by imposing on the  wave
function all first class constraints ($T$, $F^a$, $\bar F_a$) and
half of the second class constraints being in involution ($G^a$ or $\bar
G_b$). We have two quantizations:
\begin{description}
    \item -- bosonic chiral  quantization, with the   wave function satisfying the
    following wave equations
    \begin{equation}\label{chir}
    T|\Psi\rangle=0\,, \qquad F^a|\Psi\rangle=0\,, \qquad \bar F_a|\Psi\rangle=0
    \,,\qquad \bar G_a|\Psi\rangle=0
    \end{equation}
    \item -- bosonic antichiral quantization with the wave function submitted to the
    following equations
    \begin{equation}\label{achir}
    T|\Psi\rangle=0\,, \qquad F^a|\Psi\rangle=0\,, \qquad \bar F_a|\Psi\rangle=0
    \,,\qquad G^a|\Psi\rangle=0\,.
    \end{equation}
\end{description}
The reducible constraints $\bar F_a$ and $\bar G_a$ are
equivalent to primary constraint $\bar D_a$; similarly the
constraints $F^a$ and are  $G^a$ equivalent to $D^a$. Therefore one can
express the wave equations~(\ref{chir}), (\ref{achir}) in other equivalent way
\begin{description}
    \item -- bosonic chiral
 quantization:
    \begin{equation}\label{chir1}
    T|\Psi\rangle=0\,, \qquad \bar D_a|\Psi\rangle=0\,, \qquad F^a|\Psi\rangle=0
    \end{equation}
    \item -- bosonic antichiral case in which wave function is subjected the
    following constraints
    \begin{equation}\label{achir1}
    T|\Psi\rangle=0\,, \qquad D^a|\Psi\rangle=0\,, \qquad \bar F_a|\Psi\rangle=0\,.
    \end{equation}
\end{description}

Let us consider chiral case~(\ref{chir1}) in more details. Using the
realization
$$
\pi^a=-i{\partial}/{\partial\psi_a}\,,\qquad
\bar\pi_a=-i{\partial}/{\partial\bar\psi^a}
$$
and the momentum-dependent wave function $\Psi(p,\psi,\bar\psi)$ one can
write down the relations~(\ref{chir1})  as follows
\begin{equation}\label{eq-D}
\bar D_a \Psi=-i[\frac{\partial}{\partial\bar\psi^a}+(\hat p
-m)_a{}^b \psi_b ] \Psi=0\,,
\end{equation}
\begin{equation}\label{eq-F}
F^a \Psi=-i\frac{\partial}{\partial\psi_b}(\hat p +m)_b{}^a
\Psi=0\,,
\end{equation}
\begin{equation}\label{eq-T}
T \Psi=(p^2 +m^2) \Psi=0\,.
\end{equation}
The equation~(\ref{eq-D}) has the general solution
\begin{equation}\label{Psi}
\Psi(p,\psi,\bar\psi) =e^{-\bar \psi(\hat
p-m)\psi}\tilde\Psi(p,\psi)
\end{equation}
where the reduced wave  function $\tilde\Psi(p,\psi)$ depends only on $\psi$, i. e. we
have the expansion
\begin{equation}\label{tPsi}
\tilde\Psi(p,\psi) =\sum_{n=0}^\infty
\psi_{a_1}\cdots\psi_{a_n}\phi^{a_1\cdots a_n}(p)\,.
\end{equation}
Due to commuting nature of spinor $\psi_{a}$ the component fields
$\phi^{a_1\cdots a_n}(p)$ are totally symmetric
\begin{equation}\label{psi}
\phi^{a_1\cdots a_n}(p)=\phi^{(a_1\cdots a_n)}(p)\,.
\end{equation}
The equations~(\ref{eq-F}) provide the Dirac equations for these fields
\begin{equation}\label{BW}
(\hat p +m)_{a_1}{}^{b}\phi^{a_1 a_2\cdots a_n}(p)=0\,.
\end{equation}
We see that the multispinorial fields~(\ref{psi}) are Bargman--Wigner fields
describing massive particles of spins $n/2$. Obviously  the Klein--Gordon
equation~(\ref{eq-T}) is the consequence of~(\ref{BW}).

\section{Conclusion}

The classical $c$-number higher spin fields (\ref{psi}--\ref{BW})
 for any spin are
mathematically correct, and provide the
 relativistic quantum--mechanical description of one--particle states
with arbitrary mass and spin (see e. g.~\cite{GS}). The concept of
bosons and fermions is related with the symmetry properties of
multiparticle states, obtained in quantum field theory by quantum
fields acting on the vacuum state. The description of higher spin
fields presented here (see~(\ref{tPsi}), (\ref{psi})) does not take
into consideration the spin--statistics theorem, however in the
framework of first--quantized one-particle classical mechanics we
need not  to specify the statistics. The transition to the proper
spin--statistic relation can be achieved in two way:
\begin{description}
    \item i) By introducing classical theory as a suitable limit
    $\hbar \rightarrow 0$ of quantized higher spin fields. In such a
    case the half--integer spin fields will have the Grassmann
    nature (we recall that fermionic quantum fields are described by
    infinite--dimensional Clifford algebras which become in the
    limit $\hbar \rightarrow 0$ an infinite--dimensional Grassmann
    algebra).
    \item ii) One can pass from one--particle wave function to the
    wave function describing
    multiparticle states by suitable symmetrization procedure
    (besides bosonic and fermionic multiparticle states one can
    introduce also parabosonic and parafermionic multiparticle
    states, with `mixed' symmetry properties).
\end{description}

The wave functions obtained in this paper if used for the
description of multiparticle states should be therefore suitably
symmetrized: one introduces symmetric products of one--particle wave
functions for integer spin fields, and totally antisymmetric products if
spin is half--integer. Such a procedure is well-known from the description of
multi--particle states in quantum mechanics. If we wish to construct
the quantum fields which generate multiparticle states from the
vacuum we should multiply the $c$-number wave functions by
respective bosonic and fermionic creation and annihilation
operators. Such a procedure for obtaining fermionic fields with
half-integer spin or helicity can be applied to $N=1$ massless case
 (Sect.~3) and $N=2$ massive case (Sect.~4), due to the presence of linear field equations.

It should be added that $c$-number massive higher spin fields have
been obtained also in other papers from different relativistic
particle models~\cite{Has,Lyah,FedZim}. We should also add that the
realizations of `bosonic' superalgebra was used in~\cite{Lecht} for
description of physical degrees of freedom of the critical open
string with $N=2$ conformal symmetry in $2+2$ dimensions. Further
one  can point out that if one introduces fields on twistor spaces
(see e. g.~\cite{PenMac,Hug}) usually they are also commutative for
any spin, or any helicity (in massless case).

\subsubsection*{Acknowledgments} The authors would like to thank E.~Ivanov for his interest
in this paper and numerous valuable comments.

\end{document}